\documentclass[aps,prl,twocolumn,groupedaddress,showpacs]{revtex4-1}
\usepackage{graphicx}
\usepackage{dcolumn}
\usepackage{bm}
\usepackage{hyperref}
\usepackage[mathlines]{lineno}

\usepackage{amsmath}

\begin{document}


\title{Quantum Renormalization Groups Based on Natural Orbitals }



\author{Rong-Qiang He}
\author{Zhong-Yi Lu}

\email{zlu@ruc.edu.cn}


\affiliation{Department of Physics, Renmin University of China, Beijing 100872, China}


\date{\today}

\begin{abstract}

We propose a new concept upon the renormalization group (RG) procedure for an interacting many-electron correlated system in the framework of natural orbitals, and formulate an algorithm for this RG approach. To demonstrate its effectiveness, we apply this new approach on a quantum cluster-impurity model with 4 impurities in comparison with the exact diagonalization method. We also find a shortcut to dramatically improving this RG algorithm. Further discussion is presented with the cluster dynamical mean-field theory and multi-impurity (orbital) Kondo problems.

\end{abstract}

%
%

\maketitle


The physics of interacting $N$-electron correlated systems is a fascinating subject with many exciting quantum emergent phenomena. However, the theoretical study of such a system remains a fundamental issue since its computational complexity grows exponentially with $N$, which makes a direct calculation, for example, the exact diagonalization (ED), numerically intractable in practice. In order to circumvent the difficulty due to the exponential scaling, several many-body approaches have been developed, among which quantum Monte Carlo (QMC) and quantum renormalization group (RG) approaches are the most important since their accuracy can be systematically improved, namely their computational errors are controllable. However, in practice, the notorious sign problem drives the QMC back to the exponential scaling except for some special cases \cite{qmc}.

The RG approaches basically include numerical RG (NRG) and density matrix RG (DMRG) methods. The NRG was invented by Wilson to solve the single-impurity Kondo model\cite{nrg}, while the DMRG is a generalization of the NRG from energy space to real space\cite{dmrg}. The underlying idea for an RG is that we attempt to project the original full Hilbert space into a sufficiently small subspace so that we can construct a numerically tractable effective Hamiltonian whose ground state or low-energy states well approximate those of the original Hamiltonian. Thus in an RG we target the ground state and low energy properties rather than the whole energy spectrum. The essence that an RG works is that there exists such a structured subspace to approximately accommodate the ground state or targeted states that we can design iterative projections to capture this subspace according to its structure. Such subspaces in the NRG and DMRG are realized in the energy space and real space by iterative projections consisting of the lowest-energy eigenstates and the eigenvectors with the largest eigenvalues of the reduced density matrix, respectively. Both NRG and DMRG were a dramatic breakthrough in quantum many-body physics. However, the NRG works on quantum impurity problems with impurities (orbitals) no more than two, while the DMRG works basically in one-dimensional (1D) quantum systems, attributed to the topological features of their respective subspace structures.

In this Letter, we present a new quantum RG approach formulated in terms of natural orbitals, in which a structured subspace is constructed by iterative projections composed of active natural orbitals. By using this new approach, we can accurately solve general quantum impurity problems with impurities (orbitals) much over two, and with polynomial cost in the degrees of freedom of electron bath. The new approach is nonperturbative and works in the whole coupling regime with controllable errors. As we show next, the natural orbitals framework can provide an important perspective to the quantum correlation effect and then a new conceptual basis for RG approaches, in which many-body wave functions are quantitatively examined by using the natural orbitals.

Practically, we construct an $N$-electron wave function by a linear combination of $N$-electron basis functions, which are represented by Slater determinants consisting of $N$ spin-orbitals from an orthonormal set of one-electron wave functions (spin-orbitals). This is also the underlying basis for the second quantization representation. For simplicity, consider a system with $N$ interacting spinless fermions, in which $\hat{c}_{i}^{\dagger}$ ($\hat{c}_i)$ is the operator of creating (annihilating) one fermion at the $i$-th orbital and $|i\rangle=\hat{c}_i^{\dagger}|vac\rangle$, $i=1, 2, \cdots, n$, forming a complete set of orbitals. Then for a normalized $N$-fermion wave function $|\Psi\rangle$, we define the single-particle density matrix $D$ by its elements $D_{ij}\equiv\langle\Psi|\hat{c}_i^{\dagger}\hat{c}_j|\Psi\rangle$. Now we take a transformation from the orbital (site) representation $\{\hat{c}_i\}$ into a general orbital representation $\{\hat{d}_g\}$ by
\begin{equation}\label{eq_b}
(\hat{d}_1,~\hat{d}_2,~\cdots,~\hat{d}_n)=(\hat{c}_1,~\hat{c}_2,~\cdots,~\hat{c}_n)U,
\end{equation}
where $U$ is an $n\times n$ unitary matrix with elements $U_{ig}$, and $\hat{d}_g$ is the annihilation operator of the $g$-th orbital. For the $g$-th orbital, we expand $|\Psi\rangle$ as
\begin{equation}\label{eq_c}
|\Psi\rangle = \sum_{i} h_i^{(0)}|\phi_i^{0g}\rangle + \sum_j h_j^{(1)}|\phi_j^{1g}\rangle ,
\end{equation}
where $|\phi_i^{0g}\rangle$ is a Slater determinant composed of $N$ general orbitals but excluding the $g$-th orbital, namely the $g$-th orbital being unoccupied, $|\phi_j^{1g}\rangle$ a Slater determinant including the $g$-th orbital, namely occupied, and $\langle\phi_i^p|\phi_j^q\rangle =\delta_{ij}\delta_{pq}$. We introduce the expectation value $n^d_g$ of the occupation number on the $g$-th orbital, namely
$n^d_g\equiv\langle\Psi|\hat{d}_g^{\dagger}\hat{d}_g|\Psi\rangle$, and certainly $0\leq n^d_g \leq 1$. By using Eq. \eqref{eq_c}, we obtain the following important equation
\begin{equation}
n_g^d=\sum_j|h_j^{(1)}|^2 = 1 - \sum_i|h_i^{(0)}|^2 ,
\end{equation}
which quantifies the weight of each orbital in $|\Psi\rangle$.

Now we come to a crucial point. If $n^d_g$ is 0 or close to 0, $|\Psi\rangle$ will be represented or well approximated by those Slater determinants excluding the $g$-th orbital; On the contrary, if $n^d_g$ is 1 or close to 1, then including the $g$-th orbital. In both limits the degrees of freedom of the $g$-th orbital are frozen with a freezing error of $n_g^d$ and $1-n_g^d$, respectively, so it is called an inactive orbital, otherwise an active orbital. It turns out that the number of the Slater determinants in the expansion of $|\Psi\rangle$ is substantially reduced. To reduce the number of the Slater determinants as much as possible, we take the transformation (Eq. \eqref{eq_b}) in such a way that we vary $U_{ig}$ to maximize or minimize $n^d_g$ with the normalization constraint $\sum_jU_{jg}^{\ast}U_{jg}=1$ represented by the Lagrangian multiplier $\lambda_g$, noting $n^d_g=[U^{\dagger}DU]_{gg}$ by Eq. \eqref{eq_b},
\begin{equation}\label{eq_e}
\frac{\partial}{\partial U_{ig}^{\ast}}(n^d_g-\lambda_g(\sum_jU^{\ast}_{jg}U_{jg}-1)) = \sum_jD_{ij} U_{jg} - \lambda_gU_{ig}=0 ,
\end{equation}
whose solution is nothing but a complete set of the eigenvectors and eigenvalues for the single-particle density matrix $D$, namely the so-called natural orbitals with the corresponding occupation numbers\cite{book2}. Thus, among all the orbital representations, given a freezing error, the number of the Slater determinants needed to represent the wave function $|\Psi\rangle$ will be the least in the natural orbital (NO) representation.

However, a set of NOs is unknown prior to the ground state since it is defined with respect to the ground state. Hence we need to establish an iterative optimization method for finding the NOs and the ground state at the same time, namely construct a natural orbitals renormalization group (NORG). We formulate an algorithm for a general NORG as follows: (a) Take an arbitrary but complete set of orbitals. (b) Select a certain number of orbitals to form a subspace while the others remain waiting in the first iteration, otherwise replace some of the orbitals in the subspace with the largest/smallest occupancy by the same number of orbitals from those outside the subspace. (c) Construct an effective Hamiltonian matrix within the subspace and further diagonalize it to obtain the ground state. (d) Form the corresponding single-particle density matrix and diagonalize it to obtain a set of NOs and their occupancies in the subspace. (e) Return to step (b) with this new set of orbitals and iterate until convergence. Obviously the new approach will work on such a system that has a sufficient portion of natural orbitals (nearly) fully occupied or unoccupied.

As a demonstration of the effectiveness of the NORG, we have applied the NORG on a quantum impurity model in comparison with the ED method. Actually, the study of quantum impurity models has been an important topic in computational many-body physics because of the dynamical mean-field theory (DMFT) with its cluster extension (CDMFT), which maps a quantum lattice model onto a quantum impurity or a cluster of impurities coupled dynamically to a self-consistently determined bath of free electrons \cite{Georges.RMP.96,Maier.RMP.05}.

\begin{figure}
\includegraphics[width=6.0cm]{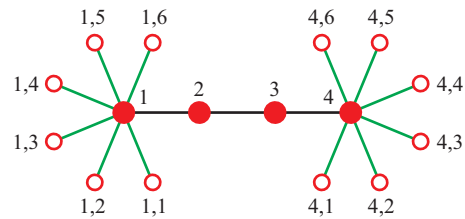}
\caption{\label{model}(Color online) Cluster-impurity model configuration with $1\times 4$ impurity sites and 12 bath sites. The filled (unfilled) circles denote the impurity (bath) sites. The links represent the electron hopping paths.}\label{fig-a}
\end{figure}

The studied cluster-impurity model is schematically shown in Figure \ref{fig-a} with 4 quantum impurities and 12 bath sites, and described by the Hamiltonian
\begin{equation}\label{eq-f}
\begin{array}{c}
\hat{H} = \hat{H}_{imp} + \hat{H}_{bath} + \hat{H}_{hyb}, \\
\hat{H}_{imp} = \sum_{ij\sigma}t_{ij} \hat{c}_{i\sigma}^\dagger \hat{c}_{j\sigma} + \sum_i U (\hat{c}^{\dagger}_{i\uparrow}\hat{c}_{i\uparrow}) (\hat{c}^{\dagger}_{i\downarrow}\hat{c}_{i\downarrow}), \\
\hat{H}_{bath} = \sum_{ik\sigma}\epsilon_{ik} \hat{c}^{\dagger}_{ik\sigma}\hat{c}_{ik\sigma},
\\
\hat{H}_{hyb} = \sum_{ik\sigma} V_{ik} \hat{c}_{i\sigma}^\dagger \hat{c}_{ik\sigma} + H.c. ,
\end{array}
\end{equation}
where $\sigma=\uparrow$ or $\downarrow$, $i$ and $j$ denote the impurity sites, and $ik$ stands for the bath sites connected with the $i$-th impurity site. To be specific, we set $t_{ij}= -t = -1$ for a pair of nearest neighbors, $t_{ii} = -U/2$ for imposing the particle-hole symmetry, and $U/t = 4$, which represents a strong correlation scenario. This model is a standard one onto which the CDMFT maps the 1D Hubbard model with $U/t=4$, in which parameters $\epsilon_{ik}$ and $V_{ik}$ are determined self-consistently (Ref. \onlinecite{modelpara}). Here we consider the half-filling case, namely one electron per site. Actually this model is close to the limit that the ED can handle since its computational complexity is about $4^{16}$.

\begin{figure}
\includegraphics[width=7.0cm]{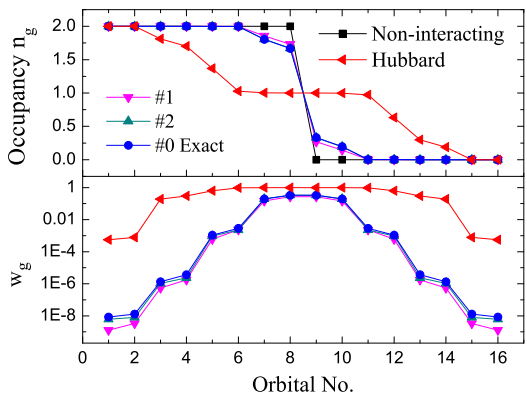}
\caption{\label{nonrm}(Color online) Calculated natural orbital occupancies. Here $w_g = min(n_g, 2-n_g)$. And \#0, 1, and 2 correspond to those in Table \ref{table1}, respectively. }\label{fig-b}
\end{figure}

In Fig. \ref{fig-b}, we show the ED-calculated NO occupancy distributions for Model \eqref{eq-f} with another two cases of $U=4$ and $U=0$ set on not only each impurity site but also each bath site, respectively, which represent three typical scenarios, namely quantum impurity model, Hubbard model, and noninteracting case, respectively. For the noninteracting, it is just a Fermi distribution, i.e., some of the NOs are doubly occupied and the others are empty. Thus the wave function is simply a Slater determinant composed of the occupied NOs. For the Hubbard case, most NOs are nearby half-filling. In contrast, for Model \eqref{eq-f} only 4 NOs are somewhat around half-filling while the others exponentially rush into double occupancy or empty. It can be shown that the number of absolutely active NOs, whose occupancy deviates well from full occupancy and empty, is about of the number of interacting impurities for a quantum impurity model. We emphasize again that this is the underlying basis for NORG working on quantum impurity models.

In practice of implementing an NORG for a quantum impurity model, we have found that a shortcut NORG is dramatically faster than a general one by about two orders of magnitude, or even more. We realize this shortcut by a restriction optimization procedure, in which a restriction is imposed on the orbital occupancy distribution according to the distribution feature so that a huge number of Slater determinants are excluded from the expansion of wave functions, and then the Hilbert space is drastically reduced. Through an iterative optimization, a set of approximate NOs is obtained. Specifically, a shortcut NORG is divided into two stages here. In Stage I we don't freeze any orbital, but apply the restriction optimization procedure to obtain a set of approximate NOs. In Stage II we freeze a certain number of orbitals according to the occupancy distribution obtained in Stage I, and then apply the restriction optimization procedure as necessary to further reduce the dimension of the subspace. In the following, we will illustrate how a shortcut NORG works on a quantum impurity model by studying Model \eqref{eq-f}.

\begin{figure}
\includegraphics[width=7.0cm]{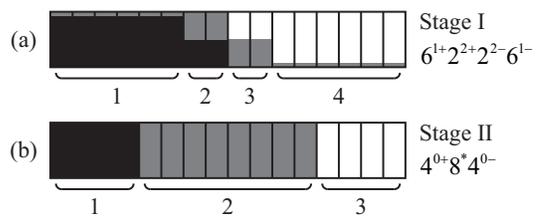}
\caption{Schematic of restriction patterns (a) and (b) successively imposed on the orbital occupancy distribution for Model \eqref{eq-f} (Fig. \ref{fig-a}). Here each small rectangle represents an orbital with at most two electrons (spin up and/or down). The total 16 orbitals are divided into a number of groups. Black (white) areas are always occupied (unoccupied), while gray areas may be occupied or unoccupied in different Slater determinants. Here a restriction pattern is represented by a series of $m^{l\pm}$ in sequence from left to right, where $m$ means $m$ orbitals in an orbital group, $l\pm$ means the maximally allowed number of holes or electrons in reference to full occupancy and empty, respectively. In addition, $*$ or $0\pm$ in a superscript means no restriction or freezing orbitals, respectively.}\label{fig-c}
\end{figure}

\begin{table*}
\caption{As a benchmark test, the shortcut NORG-calculated results in comparison with the ED-calculated ones for Model \eqref{eq-f} (Fig. \ref{fig-a}). $E_0$ stands for ground state energy and `Rel. Err.' for relative error $|(E_0-E_{\textrm{exact}})/E_{\textrm{exact}}|$ with $E_{\textrm{exact}}$ given by the ED. `GF Err.' means overall average relative error in the Green's functions. IOs stands for initial orbitals (`Site' means the original orbital representation.) `Space Size' means dimension of effective Hilbert space. The freezing error $ \epsilon_f = \sum_{g \in frozen} w_g$, where $w_g = min(n_g, 2-n_g)$. The restriction pattern is explained in Fig. \ref{fig-c}. Note that \#1 (\#2) is implemented in Stage I, followed by \#3, 4, 5, or 6 (\#7) in Stage II with initial orbitals from \#1 (\#2) (marked with `1' (`2') in IOs column). The Green's functions were calculated by the multi-targeting procedure.}\label{table1}
\begin{ruledtabular}
\begin{tabular}{r|ccr|ccc|cc}
\textrm{\#} & \textrm{Restriction Pattern} & \textrm{IOs} & \textrm{Space Size} & $E_0$ & \textrm{Rel. Err.} & $\epsilon_f$ & \textrm{GF Err.} & $\epsilon_f$ \\ \hline
0	&	$	16^{*}	$ (ED, exact)	&	Site	&	165636900	&	-34.18861144 	&	0	&	 0	&	0	&	0	\\ \hline
1	&	$	6^{1+}2^{2+}0^{*}2^{2-}6^{1-}	$	&	Site	&	3123	&	-34.06338402 	&	 3.66E-3	&	0	&	2.89E-1	&	0	\\
2	&	$	5^{1+}2^{2+}2^{*}2^{2-}5^{1-}	$	&	Site	&	21976	&	-34.18679818 	&	 5.30E-5	&	0	&	3.07E-2	&	0	\\\hline
3	&	$	4^{0+}8^{*}4^{0-}	$	&	1	&	4900	&	-34.18855985 	&	1.51E-6	&	 4.59E-6	&	2.86E-3	&	9.88E-4	\\
4	&	$	3^{0+}10^{*}3^{0-}	$	&	1	&	63504	&	-34.18859550 	&	4.66E-7	&	 1.11E-6	&	8.02E-4	&	2.07E-4	\\
5	&	$	2^{0+}12^{*}2^{0-}	$	&	1	&	853776	&	-34.18861095 	&	1.45E-8	&	 9.18E-9	&	1.96E-4	&	3.62E-5	\\
6	&	$	1^{0+}14^{*}1^{0-}	$	&	1	&	11778624	&	-34.18861123 	&	6.28E-9	&	 2.57E-9	&	2.61E-5	&	8.54E-7	\\
7	&	$	4^{0+}8^{*}4^{0-}	$	&	2	&	4900	&	-34.18857010 	&	1.21E-6	&	 6.88E-6	&	3.29E-3	&	8.89E-4	\\
\end{tabular}
\end{ruledtabular}
\end{table*}

Figure \ref{fig-c} schematically shows how a restriction is imposed on the orbital occupancy distribution for Model \eqref{eq-f}. In Stage I all 16 orbitals are divided into 4 groups, namely Group 1, 2, 3, and 4 (see Fig. \ref{fig-c}(a)). And they include 6, 2, 2, and 6 orbitals, respectively. The restriction is imposed on the occupancy distribution in such a way that Group 1 and 2 maximally allow 1 and 2 holes respectively in reference to full occupancy, denoted as $6^{1+}$ and $2^{2+}$, while Group 3 and 4 maximally allow 2 and 1 electrons respectively in reference to full empty, denoted as $2^{2-}$ and $6^{1-}$. Thus such a restriction pattern is denoted as $6^{1+}2^{2+}2^{2-}6^{1-}$. The algorithm in this stage is the same as the above one except for step (b), which is now reformulated as: Impose a proper restriction on the orbital occupancy distribution. In Stage II all 16 orbitals are divided into 3 groups. There are now 4, 8, and 4 orbitals in Group 1, 2, and 3, respectively (see Fig. \ref{fig-c}(b)). Based on the occupancy distribution obtained from stage I, Group 1 and 3 are nearly in full occupancy and empty, respectively. Both are thus frozen, denoted as $4^{0+}$ and $4^{0-}$, respectively. Group 2 is active without any restriction since the resulting subspace is small enough in this case, denoted as $8^{*}$ ($*$ means no restriction). Actually, the freezing of orbitals here can be considered as a special restriction pattern, thus denoted as $4^{0+}8^{*}4^{0-}$. The algorithm in this stage is also the same as the above one again except for step (b), which is reformulated as: Freeze those orbitals whose occupancies satisfying a pre-set freezing error and then impose a restriction as necessary.

The Green's function is a basic quantity to describe the dynamical properties of a system and also a central quantity in the (C)DMFT \cite{Georges.RMP.96,Maier.RMP.05}. Here an imaginary-frequency local Green's function matrix $G_{jk}(i\omega)$ ($j$ and $k$ being impurity-site indices) at zero temperature was studied for Model \eqref{eq-f}. In the NORG, similar to the DMRG, both excited state wave functions and Green's functions can be calculated by using the multitargeting procedure \cite{cv}, in which several targeted states are together incorporated in the calculation of the single-particle density matrix. The details will be published elsewhere.

\begin{figure}
\includegraphics[width=7.0cm]{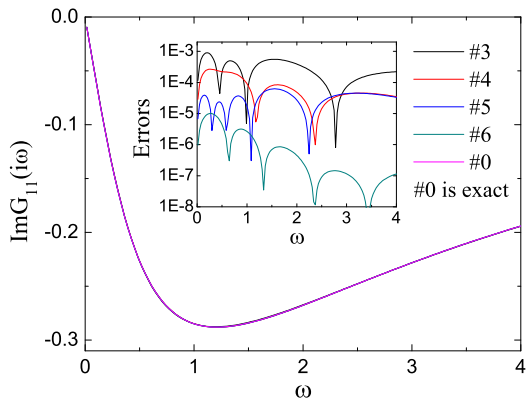}
\caption{(Color online) Green's functions calculated by the shortcut NORG with different restriction patterns (\#3 to 6) and the ED (\#0), respectively. Their errors are shown in the inset. $G_{11}(i\omega)$ is selected as a representative. Note $ReG_{11}(i\omega)=0$ due to the particle-hole symmetry. Here \#0 and 3 to 6 correspond to those in Table \ref{table1}, respectively.}\label{fig-d}
\end{figure}

Table \ref{table1} reports results for the shortcut NORG-calculated ground state energies and Green's functions of Model \eqref{eq-f} in comparison with the ED-calculated ones. The energy-converging tolerance is set to 11 digits. Although the subspace size in the shortcut NORG calculation corresponding to Fig. \ref{fig-c} is just about 1/30000 of that in the ED calculation, the energy accuracy achieved by the shortcut NORG can be as high as 6 digits (\#3 in Table \ref{table1}). Actually, when the number of the freezing orbitals decreases, the energy accuracy consistently increases, and all are excellently consistent with the corresponding freezing errors (\#3 to 6 in Table \ref{table1}). This demonstrates that a freezing error sufficiently determines the energy accuracy that an NORG will achieve. On the other hand, there are several restriction patterns available in Stage I, for instance, the restriction pattern \#2 in Table \ref{table1} is another independent implementation. They all yield good enough approximate NOs (see Fig. \ref{fig-b}) and result in almost the same accuracy in the final results (\#3 and 7 in Table \ref{table1}). The situation is the same with the Green's functions, as reported in Table \ref{table1}.

Figure \ref{fig-d} shows that the ED-calculated Green's function is accurately recovered to invisible difference by the shortcut NORG with different restriction patterns respectively (\#3 to 6 in Table \ref{table1}). Only with logarithmic vertical axis in the inset can it be shown that the differences vary with $\omega$.

Applied on a quantum impurity model, the NORG is found to take a polynomial cost ($O(N_{bath}^4)$) in the number of electron bath sites ($N_{bath}$). It turns out that dozens (even hundreds) of bath sites can be dealt with in practice \cite{suppl}. In addition, the NORG works in a Hilbert space constructed from a set of natural orbitals, thus it can work on a quantum impurity model with any lattice topological structure, unlike the NRG and DMRG.

For the (C)DMFT, the ED applied as an impurity solver shows several prominent advantages: being non-perturbative, zero-temperature reachable, and free of statistical errors, the sign problem, and the ill-posed numerical analytic continuation from imaginary frequency to real frequency to obtain real-frequency dynamics \cite{Georges.RMP.96}. Nonetheless the ED is not an optimal impurity solver in practice since its computational complexity grows exponentially with the number of not only impurity sites but also bath sites. This fundamentally limits the number of bath sites. It turns out that the dynamical mean field, represented by a set of discrete bath sites, cannot be accurately described, although the description exponentially approaches accurate with the number of bath sites \cite{ed94}. By using the NORG, we can drastically increase the number of bath sites so as to realize an accurate description, and meanwhile retain all the ED's advantages. To exemplify this aspect, we present results from the shortcut NORG-calculations for a cluster-impurity model with $2\times 2$ impurity sites and 60 bath sites in Ref. \onlinecite{suppl}.

As a major approach for studying the Kondo problems, there have been two long-standing open issues for the NRG to study multi-impurity (orbital) Kondo problems \cite{nrg1}. First, the discretization of the continuous conduction bands give rises to a lattice model, whose topological structure makes the NRG renormalization process dramatically inefficient. In contrast, the effectiveness of the NORG is basically irrespective of a model's topological structure. Second, the NRG has a poor resolution at high frequencies for dynamical quantities. In comparison, by the multitargeting procedure, the NORG, like the DMRG \cite{cv}, may treat all energy scales on equal footing.

To conclude, the unique features make the NORG naturally appropriate for studying the ground state and low-energy properties of a quantum cluster-impurity model. This will provide invaluable help in the study of the cluster dynamical mean-field theory and multi-impurity (orbital) Kondo problems.

\begin{acknowledgments}

This work is supported by National Natural Science Foundation of China (Grant Nos. 91121008 and 11190024) and National Program for Basic Research of MOST of China (Grant No. 2011CBA00112). Computational resources were provided by the Physical Laboratory of High Performance Computing in RUC.

\end{acknowledgments}


\begin{thebibliography}{}
\bibitem{modelpara} Here $\epsilon_{1k} = \epsilon_{4k}$ and $V_{1k} = V_{4k}$ due to the mirror symmetry imposed. Moreover, $\epsilon_{11} = -\epsilon_{16}$, $\epsilon_{12} = -\epsilon_{15}$, $\epsilon_{13} = -\epsilon_{14}$, $V_{11} = V_{16}$, $V_{12} = V_{15}$, and $V_{13} = V_{14}$ due to the particle-hole symmetry imposed. Then it was obtained by the exact diagonalization CDMFT that $\epsilon_{11} = 0.793777912053932$, $\epsilon_{12} = 3.60028154236076$, $\epsilon_{13} = 1.52096110386272$, $V_{11} = 0.341322224832651$, $V_{12} = 0.366149769663769$, and $V_{13} = 0.499433146276558$.
\end{thebibliography}
\end{document}



\title{Quantum Renormalization Groups Based on Natural Orbitals }



\author{Rong-Qiang He}
\author{Zhong-Yi Lu}


\affiliation{Department of Physics, Renmin University of China, Beijing 100872, China}


\date{\today}
\noindent {\bf Supplementary Material for}
\maketitle

\renewcommand{\theequation}{S\arabic{equation}}
\renewcommand{\thetable}{S\arabic{table}}
\renewcommand{\thefigure}{S\arabic{figure}}

In this Supplementary Material, the results from the shortcut NORG calculations for a cluster-impurity model with $2\times 2$ impurity sites and 60 bath sites are presented to show the practical performance of the shortcut NORG. Then a comparison between NORG and DMRG is made to help to further understand NORG. The notations and conventions follow what the main text has made unless otherwise stated.

\section{Another application example of the shortcut NORG}

\begin{figure}[b]
\includegraphics[width=0.86\columnwidth]{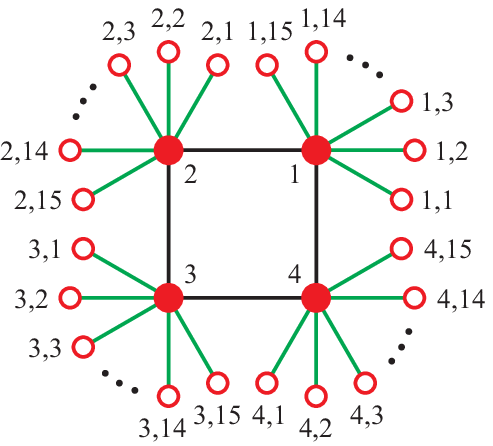}
\caption{(Color online) Visualization of the cluster-impurity model.}
\label{model2d}
\end{figure}

\begin{figure}[t]
\includegraphics[width=\columnwidth]{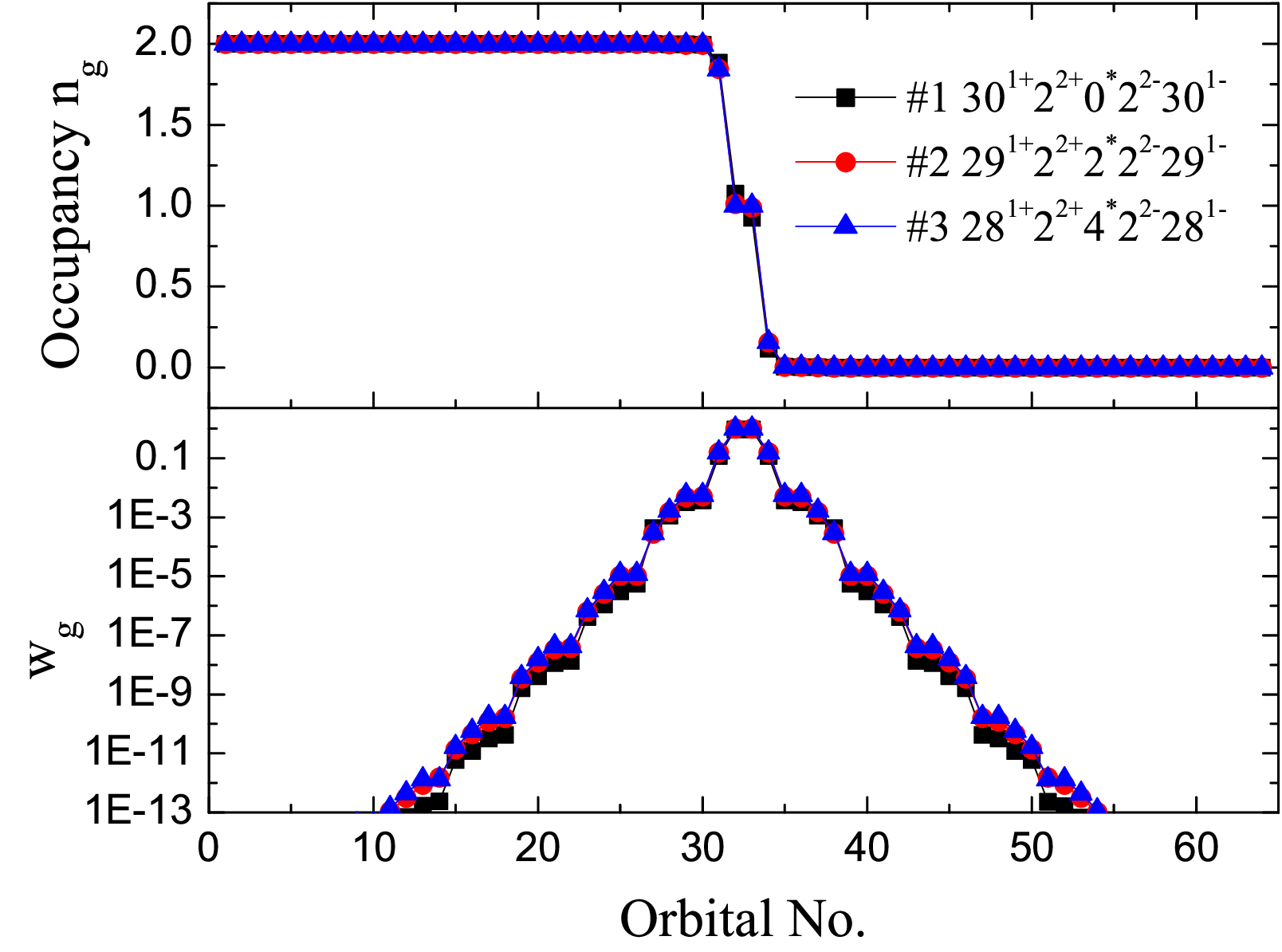}
\caption{(Color online) Occupancies of the natural orbitals obtained in Stage I of the shortcut NORG. \#1-3 correspond to those in Table \ref{gseg112d}, respectively. Here $w_g$ with values less than the machine precision don't show up in the lower panel of the figure. From \#1, \#2, to \#3, the effective Hilbert space is getting systematically larger and larger, and $w_g$ converges fast.}
\label{NO2D4}
\end{figure}

The model is schematically shown in Figure \ref{model2d}, $2\times 2$ impurity sites with 60 bath sites. $t = 1$ and $U/t = 4$. The other parameters of the model are listed in Table \ref{para2d}. The results are summarized in Figure \ref{NO2D4} and Table \ref{gseg112d}. The calculation procedure is slightly different from that in the main text, which is manifested in Table \ref{gseg112d}. In Stage II, 32 orbitals with the smallest $w_g$ determined in Stage I are frozen, while the remaining orbitals are subject to a looser restriction than that in Stage I and are further optimized. After Stage II, another stage follows, i.e., Stage III, in which 12 more orbitals with the smallest $w_g$ are frozen, while the remaining orbitals are subject to a looser restriction than that in Stage II and we get more accurate physical quantities.

\begin{table}[b]
\caption{Some of the parameters of the model. $\epsilon_{1k} = \epsilon_{2k} = \epsilon_{3k} = \epsilon_{4k}$ and $V_{1k} = V_{2k} = V_{3k} = V_{4k}$ ($k = 1, 2, ..., 15$) due to the four-fold rotational symmetry. $\epsilon_{1k} = -\epsilon_{1,16-k}$ and $V_{1k} = V_{1,16-k}$ ($k = 1, 2, ..., 8$) due to the particle-hole symmetry imposed.}
\label{para2d}
\begin{ruledtabular}
\begin{tabular}{ccc}
$k$ & $\epsilon_{1k}$ & $V_{1k}$  \\ \hline
1	&	1.09432050353993E+00	&	-3.25861700524933E-01	\\
2	&	-8.12129452137460E-01	&	-2.53572949534344E-01	\\
3	&	2.69973133535642E+00	&	-2.98856870563701E-01	\\
4	&	-6.27254332635770E+00	&	8.40922667355125E-02	\\
5	&	-6.45481830221366E-01	&	1.34235217456270E-01	\\
6	&	4.02420994365745E+00	&	2.45919037006646E-01	\\
7	&	-1.63387758294565E+00	&	3.93233653967598E-01	\\
8	& 	0.00000000000000E+00	&	-3.76505701495740E-06	\\
\end{tabular}
\end{ruledtabular}
\end{table}

\begin{table*}[t]
\caption{The results for the shortcut NORG calculations. \#1-3 are independently implemented in Stage I, \#4-7 in Stage II, and \#8-11 in Stage III. The result of \#11 is used as a reference to estimate the errors of the results of \#1-9 (see Table \ref{roptest}). Some errors of the Green's functions are not given because of extra computational cost. The errors of the final results (Stage III, i.e., \#8-11) include those introduced by the restriction optimization procedure in Stage III (using, for example, in \#8 and \#9 restriction pattern $22^{0+}8^{2+}4^{*}8^{2-}22^{0-}$ instead of $22^{0+}20^{*}22^{0-}$ of which the calculation can not be implemented) and those introduced by the successive orbital freezing in Stages II and III (which are the genuine errors due to the NORG renormalization processes). Both of them are controllable and estimable. See the main text and Table \ref{roptest}, respectively. For \#8, as a final shortcut NORG result, the ground state energy has an error of 3.91E-7 and the Green's function has an error of 6.52E-4, whereas the largest effective Hilbert space involved in the calculations is of dimension 427624 while that in a typical ED calculation is 3.36E+36, which shows the great power of the shortcut NORG.}
\label{gseg112d}
\begin{ruledtabular}
\begin{tabular}{r|ccr|ccc|cc}
\# & Restriction Pattern & IOs & Space Size & $E_0$ & \textrm{Rel. Err.}& $\epsilon_f$ & GF Err. & $\epsilon_f$ \\ \hline
1	&	$	30^{1+}2^{2+}0^{*}2^{2-}30^{1-}	$	&	Site	&	71667	&	-148.5323438591 	 &	1.11E-3	&	0	&	--	&	0	\\	
2	&	$	29^{1+}2^{2+}2^{*}2^{2-}29^{1-}	$	&	Site	&	641368	&	-148.6909791787 	 &	4.60E-5	&	0	&	--	&	0	\\	
3	&	$	28^{1+}2^{2+}4^{*}2^{2-}28^{1-}	$	&	Site	&	7186598	&	-148.6975369976 	 &	1.94E-6	&	0	&	--	&	0	\\	\hline
4	&	$	16^{0+}15^{2+}2^{*}15^{2-}16^{0-}	$	&	1	&	427624	&	-148.6860293001 	 &	7.93E-5	&	3.71E-11	&	--	&	3.75E-8	\\	
5	&	$	16^{0+}15^{2+}2^{*}15^{2-}16^{0-}	$	&	2	&	427624	&	-148.6860292991 	 &	7.93E-5	&	1.25E-10	&	--	&	1.06E-7	\\	
6	&	$	16^{0+}15^{2+}2^{*}15^{2-}16^{0-}	$	&	3	&	427624	&	-148.6860292987 	 &	7.93E-5	&	1.57E-10	&	--	&	7.24E-8	\\	
7	&	$	16^{0+}14^{2+}4^{*}14^{2-}16^{0-}	$	&	3	&	3573606	&	-148.6977692706 	 &	3.83E-7	&	1.57E-10	&	--	&	7.24E-8	\\	 \hline
8	&	$	22^{0+}8^{2+}4^{*}8^{2-}22^{0-}	$	&	4	&	404580	&	-148.6977680385 	&	 3.91E-7	&	4.79E-8	&	6.52E-4	&	8.77E-6	\\	
9	&	$	22^{0+}8^{2+}4^{*}8^{2-}22^{0-}	$	&	7	&	404580	&	-148.6977680646 	&	 3.91E-7	&	9.32E-8	&	4.79E-4	&	2.21E-5	\\	
10	&	$	22^{0+}8^{3+}4^{*}8^{3-}22^{0-}	$	&	4	&	9160292	&	-148.6978260933 	&	 --	&	4.79E-8	&	--	&	8.77E-6	\\	
11	&	$	22^{0+}8^{3+}4^{*}8^{3-}22^{0-}	$	&	7	&	9160292	&	-148.6978261476 	&	 --	&	9.32E-8	&	--	&	2.21E-5	\\	
\end{tabular}
\end{ruledtabular}
\end{table*}

\begin{table*}[t]
\caption{A benchmark test for the restriction optimization procedure. The initial orbitals for the three calculations are the resulting natural orbitals of \#4 in Table \ref{gseg112d}. 50 orbitals are frozen. For the remaining orbitals, \#3 is an exact result while \#1 and \#2 are approximate. The errors of the results of \#1 and \#2 are estimated with reference to the result of \#3. \#2 is two orders of magnitude more accurate than \#1. So one can estimate the error of the result of \#1 by considering that of \#2 as a reference.}
\label{roptest}
\begin{ruledtabular}
\begin{tabular}{rcrccc}
\# & Restriction Pattern & Space Size & $E_0$ & \textrm{Rel. Err.} & GF Err. \\ \hline
1	&	$	25^{0+}5^{2+}4^{*}5^{2-}25^{0-}	$	&	67656	&	-148.69749932305 	&	3.90E-7	 &	4.69E-4	\\
2	&	$	25^{0+}5^{3+}4^{*}5^{3-}25^{0-}	$	&	582456	&	-148.69755653670 	&	4.83E-9	 &	6.03E-6	\\
3	&	$	25^{0+}14^{*}25^{0-}	$	&	11778624	&	-148.69755725526 	&	0	&	0	 \\
\end{tabular}
\end{ruledtabular}
\end{table*}

\begin{table*}[t]
\caption{Summary of the comparison between NORG and DMRG.}
\label{cmpdmrg}
\begin{tabular}{|c|c|c|}
\hline
Method & NORG & DMRG \\ \hline
Central concept            & Single-particle density matrix & Reduced density matrix \\
Characteristic quantity    & Natural-orbital occupancy      & Entanglement spectrum \\
Variational state          & Orbital-frozen state           & Matrix product state \\
Favorite interaction form  & Sparse interactions            & Short-range interactions \\
Favorite spatial structure & Unrestricted                   & One dimensional chain \\
Many-body feature          & Sparse correlation             & Low entanglement \\
\hline
\end{tabular}
\end{table*}

\clearpage

\section{A comparison between NORG and DMRG}

It is well-known that DMRG works best on 1D systems with short-range interactions. The underlying reason is that such systems often possess low entanglement. Then the entanglement spectrum, namely the eigenvalues of the reduced density matrix, shows a rapid-decay (often exponential-decay) feature, which allows an efficient truncation of many contribution-negligible states and a good description of the ground state by a matrix product state. In comparison, NORG works best on fermionic systems with sparse interactions, namely a small number of interacting sites and a large number of non-interacting sites (quantum multisite and/or multiorbital impurity models are excellent examples). The underlying reason is that such systems often possess `sparse correlation', i.e., the ground state wave function can be expanded by a small number of Slater determinants compared to the dimension of the complete Hilbert space of the systems. (See below for the detailed explanation of the concept of `sparse correlation'.) Then the natural-orbital occupancy, namely the eigenvalues of the single-particle density matrix, shows a rapid-decay (often exponential-decay) feature, i.e., the occupancy quickly approaches full occupancy or empty, which allows an efficient truncation of many contribution-negligible orbitals and a good description of the ground state by an orbital-frozen state. These discussions are summarized in Table \ref{cmpdmrg}.

The NORG described in the left column in the second page of the main text, as a straightforward realization of the spirit above, is an iterative RG. In each iteration some orbitals are added in. After a diagonalization, some orbitals with small contribution are frozen out. In some sense, this process looks like the finite-system DMRG, in which in each iteration the full degrees of freedom of few sites are added in and after a diagonalization some degrees of freedom (states) with small contribution are frozen out (truncated out). This straightforwardly realized NORG works indeed, but not so efficiently as expected. Since we have been aware of the above spirit and reality allowing the existence of an efficient RG approach on fermionic systems with sparse interactions, resembling the case that DMRG works on 1D quantum systems with short-range interactions, and been aware of that the RG procedure of DMRG is merely to find a small structured subspace to accommodate a specific variational wave function (i.e. the matrix product state) with the ability describing accurately the ground state, we don't have to confine our mind to strict iterative RG. Then the shortcut NORG comes out, in which more sufficient preparatory work (finding a set of relatively good natural orbitals) is done before a tremendous renormalization (freezing many orbitals) takes place. Such a shortcut NORG is so efficient that sometimes only one truncation is needed in solving a system (like the one in the main text), and sometimes more than one (like the one in the Supplementary Material).

Here we explain the concept of `sparse correlation' for fermionic systems qualitatively in terms of Slater determinants. Consider a many-body wave function. If it can be expressed exactly by only one Slater determinant, it possesses no correlation, which means that the Hartree-Fock theory provides an exact description to it. If it can be well approximated by a single Slater determinant, it possesses weak correlation, which means that the Hartree-Fock theory provides a good but approximate description to it. Otherwise, it possesses strong correlation. Then we ask the question how many Slater determinants are needed to (approximately) well describe a wave function possessing strong correlation. A problem that one encounters is that the number of the Slater determinants may depend on the orbital set that one chooses. Therefore, we only consider the minimum possible number of the Slater determinants needed to (approximately) well describe the wave function. If the logarithm of the number is substantially smaller than the number of the single-particle degrees of freedom (orbitals or sites) of the wave function, the wave function possesses sparse correlation. Otherwise the wave function possesses dense correlation. Among all the orbital representations, the minimum possible number appears when the wave function is expanded in the natural-orbital representation, as shown in our manuscript. Through practice we now know that the logarithm of the minimum possible number of the Slater determinants needed to (approximately) well describe the ground state of a general fermionic system is roughly proportional to the number of its interacting sites. If all the sites in a system are interacting, a Hubbard model for example, then the logarithm of the number is roughly proportional to the size of the system. This means that quantum impurity models possess sparse correlation while Hubbard-like models possess dense correlation.